\documentclass[10pt]{article}

\usepackage{amsmath}
\usepackage{amssymb}
\usepackage{graphicx}
\usepackage{cite}
\usepackage{color}
\usepackage{subfigure} 
\usepackage{epstopdf}
\usepackage[hidelinks]{hyperref}

\usepackage[labelfont=bf,labelsep=period,justification=raggedright]{caption}

\makeatletter
\renewcommand{\@biblabel}[1]{\quad#1.}
\makeatother

\date{}

\begin{document}

% Title must be 150 characters or less
\begin{flushleft}
{\Large
\textbf{Mapping the  Space of Genomic Signatures}
}
\\
\vspace*{5mm}
Lila Kari$^{1,\ast}$, 
Kathleen A. Hill$^{1, 2}$, 
Abu S. Sayem$^1$,
Rallis Karamichalis$^1$,
Nathaniel Bryans$^1$,
Katelyn Davis$^2$,
Nikesh S. Dattani$^3$
\\
\vspace*{2mm}
\bf{1} Department of Computer Science, University of Western Ontario, London, ON, N6A 5B7, Canada
\\
\bf{2} Department of Biology, University of Western Ontario, London, ON, N6A 5B7, Canada
%\\
%\bf{3} Microsoft Corporation, Redmond, WA,  98052, United States of America
\\
\bf{3} Physical and Theoretical Chemistry Laboratory, Department of Chemistry, Oxford University, Oxford, OX1 3QZ, United Kingdom
\\
$\ast$ E-mail: lila.kari@uwo.ca
\end{flushleft}

% Please keep the abstract between 250 and 300 words
\section*{Abstract}

We propose a computational method to measure and visualize interrelationships among any number of DNA sequences allowing, for example, the examination of hundreds or thousands of complete mitochondrial genomes.  An ``image distance'' is computed for each pair of  graphical representations of DNA sequences, and the distances are visualized as a Molecular Distance Map: Each point on the map represents a DNA sequence, and the spatial proximity between any two points reflects the degree of structural similarity between the corresponding sequences. The graphical representation of DNA sequences utilized, Chaos Game Representation (CGR),  is genome- and species-specific and can thus act as a genomic signature. Consequently, Molecular Distance Maps could inform
species identification, taxonomic classifications and, to a certain extent, evolutionary history.  The image distance employed, Structural Dissimilarity Index (DSSIM), implicitly compares the occurrences of oligomers of length up to $k$ (herein $k=9$) in DNA sequences. We computed DSSIM distances for more than 5 million pairs of complete mitochondrial genomes, and used Multi-Dimensional Scaling (MDS) to obtain Molecular Distance Maps that visually display the sequence relatedness in various subsets, at different taxonomic levels. 

This general-purpose method does not require DNA sequence homology and can thus be used to compare similar or vastly different DNA sequences, genomic or computer-generated, of the same or different lengths.  We illustrate potential uses of this approach by applying it
 to several taxonomic subsets: phylum Vertebrata, (super)kingdom Protista, classes Amphibia-Insecta-Mammalia, class Amphibia, and order Primates.  This analysis of an extensive dataset  confirms that the  oligomer composition of full mtDNA sequences can be a source of taxonomic information. This method also correctly finds the mtDNA sequences most closely related to that of the anatomically modern human (the Neanderthal, the Denisovan, and the chimp), and that the sequence most different from it belongs to a cucumber.

% Please keep the Author Summary between 150 and 200 words
% Use first person. PLoS ONE authors please skip this step. 
% Author Summary not valid for PLoS ONE submissions.   
%\section*{Author Summary}

\section*{Introduction}

Every year biologists discover and classify thousands of new species, with  an average of  18,000 new species named  each year  since   1940 \cite{IISE2012}, \cite{NationalGeographic2014}. Other findings, \cite{Mora2011}, suggest that as many as 86\% of existing species on Earth and 91\% of species in the oceans have  not yet been  classified and catalogued. In the absence of a universal quantitative method to identify species' relationships,  information for species classification has to be gleaned  and combined  from several sources, morphological, sequence-alignment-based phylogenetic anaylsis, and non-alignment-based molecular information.
 
 We propose a computational process that  outputs, for any given dataset of DNA sequences, a  concurrent display of  the structural similarities among all sequences in the dataset. This is obtained by first computing an ``image distance'' for each pair of  graphical representations of DNA sequences, and then  visualizing the resulting interrelationships in  a two-dimensional plane. The result of applying this method to a  collection of DNA sequences is  an easily interpretable {\it Molecular Distance Map}  wherein sequences are represented by points in a common Euclidean plane,  and the spatial distance between any two points reflects the differences in  their subsequence composition.
 
 The graphical representation we use is {\it Chaos Game Representation} (CGR) of DNA sequences, \cite{Jeffrey1990,Jeffrey1992},   that  simultaneously displays all subsequence frequencies of a given DNA sequence as a visual pattern. CGR has a remarkable ability to differentiate between genetic sequences belonging to different species, and has thus been proposed as a {\it genomic signature}.  Due to this characteristic, a Molecular Distance Map of a  collection of genetic sequences   may allow inferrences of relationships between the corresponding species.

Concretely, to compute and visually display relationships in a given set $S = \{s_1, s_2, ..., s_n \}$ of $n$ DNA sequences, we propose the following computational process:

\noindent

(i) {\em Chaos Game Representation} (CGR), to graphically  represent all subsequences of a DNA sequence $s_i$, $1\leq i\leq n$, as pixels of one image, denoted by $c_i$;

(ii)  {\em Structural Dissimilarity Index} (DSSIM), an ``image-distance'' measure, to compute the pairwise distances $\Delta (i, j)$, $1\leq i, j \leq n$, for each  pair of CGR images $(c_i, c_j)$, and to produce a  distance matrix;

(iii) {\em Multi-Dimensional Scaling} (MDS), an information visualization technique that  takes as input the distance matrix and outputs  a Molecular Distance Map in 2D, wherein each plotted point $p_i$ with coordinates $(x_i, y_i)$ represents the DNA sequence $s_i$ whose CGR image is $c_i$. The position of the point $p_i$ in the map, relative to all the other points $p_j$,  reflects the  distances between the DNA sequence $s_i$  and the other  DNA sequences $s_j$ in the dataset.

We apply this method to analyze and visualize several different taxonomic subsets of a dataset of 3,176 complete mtDNA sequences:  phylum Vertebrata, (super)kingdom Protista, classes Amphibia-Insecta-Mammalia, class Amphibia only, and order Primates.  We illustrate  the usability of this approach by discussing, e.g.,  the placement  of  the genus  {\it Polypterus} within  phylum Vertebrata, of   the unclassified organism  {\it Haemoproteus} sp. jb1.JA27 (\#1466)  within the (super)kingdom Protista, and the  placement of the family Tarsiidae within the order Primates. We also provide an interactive web tool, {\it MoD-Map}, that allows  an in-depth exploration of all Molecular Distance Maps in the paper, complete  with zoom-in features,  search options, and easily accessible additional information for each species-point  sequence.

Overall, this method groups   mtDNA sequences in correct taxonomic groups, from the kingdom level  down to the order and family  level.  These results are of interest both because of the size of the dataset analyzed and because this information was extracted from DNA sequences that normally would not be considered homologous. Our analysis confirms  that  sequence composition (presence  or absence of oligomers) contains taxonomic information that could be relevant to species identification, taxonomic classification, and  identification of large evolutionary lineages. Last but not least, the appeal of this method lies in its simplicity, robustness, and generality, whereby the exact same measuring tape is able to automatically yield meaningful measurements  between non-specific DNA  sequences of species as distant as  those of  the anatomically modern human and a cucumber,  and as close as those of the  anatomically modern human and the Neanderthal.

% Results and Discussion can be combined.

\section*{Methods}

 A CGR \cite{Jeffrey1990, Jeffrey1992} associates an image to each DNA sequence as follows. Begin with  a unit square with corners  labelled {\it A, C, G,} and {\it T}, clockwise starting from the bottom-left corner. The first point of any CGR plot is the center of the square.  To plot the CGR  corresponding to a given DNA sequence, start reading the letters  of the sequence from left to right, one by one. The  point corresponding to the first letter  is the point plotted in the middle of the segment determined by the center of the  square and the corner labelled by the first  letter. For example, if the center of the square is labelled ``O'' and the first letter of the sequence is ``A'', then the point of the plot corresponding to the first ``A'' is the point situated halfway between O and the corner A. Subsequent  letters are  plotted iteratively as the middle point between the previously-drawn-point and the  corner labelled by the letter currently being read. 
 
 CGR images of genetic DNA sequences originating from various species show  rich fractal patterns containing various motifs such as squares, parallel lines, rectangles, triangles and diagonal crosses, see, e.g.,  Figure~\ref{fig:CGR}.    CGRs  of genomic DNA sequences  have been shown to be  genome and species specific, \cite{Jeffrey1990,Jeffrey1992,Hill1992,Hill1997,Deschavanne1999, Deschavanne2000,Wang2005}. Thus,  sequences chosen from each genome as a basis for computing ``distances'' between genomes do not need to have any relation with one another from the point of view of their position or information content.  In addition, this graphical representation facilitates easy visual recognition of global string-usage characteristics:  Prominent diagonals indicate purine or pyrimidine runs, sparseness in the upper half indicates low G+C content, etc., see, e.g.,  \cite{Deschavanne1999}.

If the generated CGR image has a resolution  of $2^k\times 2^k$ pixels, then every pixel represents a distinct DNA subsequence of length $k$: A pixel is black if the subsequence it represents occurs in the DNA sequence, otherwise it is white. In this paper, for the CGR images of all 3,176 complete  mtDNA  sequences in our dataset,   we used the value $k = 9$, that is, occurrences of subsequences of lengths up to 9  were being taken into consideration.  In general, a length of the DNA sequence of 4,000 bp is necessary to obtain a sharply  defined CGR, but in many cases 2,000 bp give a reasonably good approximation, \cite{Jeffrey1990}.
  %Other studies such as \cite{Deschavanne2000}   suggest that DNA sequences of lengths on the order $10^{4} $ to $10^5$ be used. 
    In our case, we  used the full length of all analyzed mtDNA sequences, which ranged from 288 bp to 1,555,935 bp, with an average of 28,000 bp.
     
Other visualizations of genetic data include the 2D rectangular walk \cite{Gates1986} and  methods similar to it  in \cite{Nandy1994},\cite{Leong1995}, vector walk  \cite{Liao2005},  cell \cite{Yao2004},    vertical vector \cite{Yu2010},  Huffman coding  \cite{Qi2011}, and  colorsquare \cite{Zhang2012} methods. Three-dimensional representations of DNA sequences include  the tetrahedron   \cite{Randic2000}, 3D-vector  \cite{Yuan2003}, and  trinucleotide curve \cite{Yu2009}  methods. Among  these visualization methods, CGR images arguably provide the most immediately comprehensible  ``signature'' of a DNA sequence and a desirable genome-specificity, \cite{Jeffrey1990,Deschavanne2000}. In  addition, the  images produced using CGR  are easy to compare, visually and computationally. Coloured versions of CGR, wherein the colour of a point corresponds  to the frequency of the corresponding oligomer in the given DNA sequence (from red for high frequency, to blue for no occurrences) have also been proposed \cite{Makula2009, Hao2000}.
  
Note that other alignment-free methods have been  used for phylogenetic analysis of DNA strings, such as  computing the Euclidean distance between frequencies of $k$-mers ($k \leq 5$) for the analysis of  125 GenBank DNA sequences from 20 bird species and the American alligator, \cite{Edwards2002}. Another study, \cite{Deschavanne2010},  analyzed  459 dsDNA bacteriophage genomes and compared them with their host genomes to infer  host-phage  relationships,  by computing Euclidean distances  between frequencies of $k$-mers  for $k=4$. In \cite{Pandit2010},  75   complete HIV genome sequences were compared using the Euclidean distance between frequencies of 6-mers ($k=6$), in order to group them in subtypes. In \cite{Pride2003},  27 microbial genomes  were analyzed  to find  implications of  4-mer frequencies  ($k=4$) on their evolutionary relationships.  In \cite{Li2004}, 20  mammalian  complete mtDNA  sequences were analyzed using  the ``similarity metric''. Our method uses a  larger dataset (3,176  complete mtDNA sequences), an ``image distance'' measure that was designed to capture structural similarities between images, as well as a value of  $k=9$. 

Structural Similarity (SSIM) index is an image similarity index  used in the context of image processing  and computer vision to compare two  images from the point of view of their structural similarities \cite{Wang2004}.  SSIM combines three parameters - luminance distortion, contrast distortion, and linear correlation - and was designed to perform similarly to the human visual system, which is highly adapted to extract  structural information. Originally, SSIM  was defined as a similarity measure $s(A,B)$ whose theoretical range between two  images $A$ and $B$ is $[-1,1]$ where a high value amounts to close relatedness. We use a related {\em DSSIM distance} $\Delta(A,B) = 1 - s(A,B) \in [0,2]$, with the distance being 0 between two identical images, 1 between e.g.\ a black image and a white image, and 2 if the two images are negatively correlated; that is, $\Delta(A,B) = 2$ if and only if every pixel of image $A$ has the inverted value of the corresponding pixel in image $B$ while both images have the same luminance (brightness). For our particular dataset of genetic CGR images, almost all (over 5 million) distances are between 0 and 1, with  only half a dozen exceptions of distances between 1 and 1.0033.

MDS  has been used for the visualization of data relatedness  based on distance matrices in various fields such as cognitive science, information science, psychometrics, marketing, ecology, social science, and  other areas of study \cite{BorgGroenen2010}. MDS takes as input a distance matrix  containing the pairwise distances between $n$ given  items and outputs a two-dimensional map wherein each item is represented by a point, and the spatial distances between points reflect the distances between the corresponding items in the distance matrix. Notable examples of molecular biology studies that used MDS are \cite{Lessa1990} (where it was used for the analysis of geographic genetic distributions of some natural populations),  \cite{Hebert2003} (where it was used to provide a graphical summary  of the distances among  CO1 genes from  various species), and \cite{Hillis2005} (where it was used to analyze and visualize relationships within collections of   phylogenetic trees).

Classical MDS, which we use in this paper, receives as input an $n\times n$ distance matrix  $(\Delta(i, j))_{1\leq i, j \leq n}$ of the pairwise distances between any two items in the set. The output of classical MDS consists of $n$ points in a $q$-dimensional space whose pairwise spatial (Euclidean)  distances are a linear function of the distances between the corresponding items in the input distance matrix. More precisely, MDS will return $n$ points $p_{1},p_{2},\ldots,p_{n}\in \mathbb{R}^{q}$ such that $d(i, j)= ||p_{i}-p_{j}||\thickapprox f(\Delta(i,j))$ for all $i,j\in \{1, \ldots, n \}$ where $d(i, j)$ is the spatial distance between  the points $p_i $ and $p_j$, and $f$ is a  function linear in $\Delta(i, j)$. Here, $q$ can be at most $n-1$ and the points are recovered from the eigenvalues and eigenvectors of the input $n\times n$ distance matrix. If we choose $q=2$ (respectively $q=3$), the result of classic MDS is  an approximation of the original $(n-1)$-dimensional space as a two- (respectively three-) dimensional map.

In this paper all Molecular Distance Maps  consist of coloured points, wherein each point represents an mtDNA sequence from the dataset. Each mtDNA sequence is assigned a unique numerical identifier retained in all analyses, e.g., \#1321 is the identifier for the {\it Homo sapiens sapiens} mitochondrial genome. The  colour assigned to a sequence-point  may however  vary from map to map, and it depends on  the taxon assigned to the point in a particular Molecular Distance Map. For example, in Figure  \ref{vertebrates} all mammalian mtDNA sequence-points  are coloured red, while in Figure  \ref{primatesonly} the red points represent  mtDNA sequences from  the primate suborder Haplorhini and the green points represent mtDNA sequences from the primate suborder Strepshirrini. For consistency, all maps  are scaled so that the $x$- and the $y$-coordinates always span  the  interval $[-1, 1]$. The formula used for scaling is $x_{\rm{sca}}  =2 \cdot (\frac{x - x_{\rm{min}}}{x_{\rm{max}} - x_{\rm{min}}}) - 1$, $y_{\rm{sca}}  =2 \cdot (\frac{y - y_{\rm{min}}}{y_{\rm{max}} - y_{\rm{min}}}) - 1$, where $x_{\rm{min}}$ and  $x_{\rm{max}}$ are the minimum and maximum of  the $x$-coordinates  of all the points in the original map, and similarly  for $y_{\rm{min}}$ and $y_{\rm{max}}$.

Each Molecular Distance Map  has some error, that is, the spatial distances $d_{i, j}$ are not exactly the same as  $f(\Delta(i,j))$. When using the same dataset, the error is in general lower for an MDS map in a higher-dimensional space. The {\it Stress-1} (Kruskal stress, \cite{Kruskal64}), is defined in our case as

$$\mbox{{\it Stress-1}} = \sigma_1 = \sqrt{\frac{\Sigma_{i < j} [f(\Delta(i,j)) - d_{i,j}]^2}{\Sigma_{i < j} d_{i,j}^2}}$$

\noindent
where  the summations extend over all the sequences considered for a given map, and $f(\Delta(i,j)) = a \times \Delta(i, j) + b$  is a linear function whose parameters $a, b \in \mathbb{R}$ are determined by linear regression for each dataset and corresponding Molecular Distance Map. A benchmark that is often used to assess MDS results  is that  {\it Stress-1} should  be in the range  $[0, 0.20]$, see \cite{Kruskal64}.

The dataset  consists of the entire  collection of complete mitochondrial DNA sequences from NCBI as of 12 July, 2012.  This dataset consists of  3,176 complete mtDNA sequences, namely  79 protists, 111 fungi, 283 plants, and 2,703 animals. This collection of mitochondrial genomes has a great breadth of species  across taxonomic categories and great depth of species coverage in certain taxonomic categories. For example, we compare sequences at every rank of taxonomy, with some pairs being different at as high as the (super)kingdom level, and some pairs of sequences being from the exact same species, as in the case of {\it Silene conica} for which our dataset contains the sequences of 140 different mitochondrial chromosomes \cite{Sloan2012}. The prokaryotic origins and  evolutionary history  of mitochondrial genomes have long been extensively studied, which will allow comparison  of our results with  known relatedness of species. Lastly, this genome dataset permits testing of both recent and deep rooted species relationships, providing fine resolution of species differences.

The creation of the datasets,  acquisition of data from NCBI's GenBank,  generation of the CGR images, calculation of the distance matrix, and  calculation of the Molecular Distance Maps using MDS, were all done (and can be tested with) the free open-source MATLAB program OpenMPM  \cite{Dattani2013}. This program makes use of an open source MATLAB program for SSIM, \cite{Wang2003},
 and MATLAB's built-in  MDS function. The  interactive  web tool  {\it MoD-Map} ({\it Mo}lecular {\it D}istance {\it Map}), \cite{MoDMap}, allows an in-depth exploration and navigation of the Molecular Distance Maps in this paper$^1$. 

\footnotetext[1]{On-line Supplemental Material includes the annotated dataset and the  DSSIM distance matrix,  and  can be found at 
\url{http://www.csd.uwo.ca/\~lila/MoDMap/}.  When using the  web tool  {\it MoD-Map}, \cite{MoDMap}, clicking on the ``Draw MoD Map'' button allows selection of any of the five maps presented in the paper, each  with   features such as zoom-in  and search  by scientific name of the species  or the NCBI accession number of its mtDNA.  On any given MoD map, clicking on a sequence-point  displays its full mtDNA  sequence information such as its unique identifier in this analysis, NCBI accession number, scientific name, common name, length of mtDNA sequence, taxonomy, CGR plot, as well as a link to the corresponding NCBI entry. Clicking on the ``From here'' and ``To here'' buttons displays the image distance between the  CGR plots of  two selected sequence-points,  as a number between 0 and 1.
}

\section*{Results and Discussion}

The Molecular Distance Maps we analyzed,  of several  different taxonomic subsets (phylum Vertebrata, (super)kingdom Protista,  classes Amphibia-Insecta-Mammalia, class Amphibia only, and order Primates), confirm that the presence or absence of oligomers in mtDNA sequences may contain  information that is relevant to taxonomic classifications.  These results are of interest both because of the large dataset considered and  because this information has been extracted from DNA sequences  that, by normal criteria, would be  considered nonhomologous. The main contributions of the paper are the following:

\begin{itemize}

  \item The use of an ``image distance'' (designed to detect structural similarities between images) to compare the graphic signatures of two DNA sequences. For any given $k$, this distance simultaneously compares the  occurrences of all  subsequences  of length up to $k$ of the two sequences. 
 In all computations of this paper we use $k=9$.    This image distance (with parameter set to $k=9$)  is highly sensitive and  succeeds to successfully group hundreds of CGRs that are visually similar, such as the ones in Figure \ref{fig:CGR}(left) and   Figure \ref{fig:CGR}(right), into correct taxonomic categories.  

\item The use of  an information visualization  technique to display the results as  easily interpretable Molecular Distance Maps, wherein the spatial position of each sequence-point in relation to all other sequence-points is quantitatively  significant.  This is augmented by an interactive web tool which allows an  in-depth exploration of the Molecular Distance Maps in this paper, with features such as zoom-in, search by scientific name or NCBI accession number, and quick access to complete  information for each of the full mtDNA sequences in the map.

\item  A method that is general-purpose, simple,  computationally efficient and scalable.  Since the compared sequences need not be homologous or of the same length, this method  can be used to  provide comparisons  among any number of  completely different DNA sequences: within the  genome of an individual, across genomes within a single species, between genomes within a taxonomic category, and across taxa. 
 
 \item The use of a  large dataset of  3,176 complete mitochondrial DNA sequences.
 
 \item An illustration of potential uses of this approach by  the discussion of several case studies such as the placement of  the genus  {\it Polypterus} within  phylum Vertebrata, of   the unclassified organism  {\it Haemoproteus} sp. jb1.JA27 (\#1466)  within the (super)kingdom Protista, and the  placement of the family Tarsiidae within the order Primates.

\end{itemize}
 
This method could  complement information obtained by  using DNA barcodes \cite{Hebert2003} and Klee diagrams \cite{Sirovich2010}, since  it is applicable to cases where barcodes may have limited effectiveness: plants and fungi for which different barcoding regions have to be used \cite{Kress2005},   \cite{Hollingsworth2009}, \cite{Schoch2011}; protists where multiple loci are generally needed to distinguish between species \cite{Hoef2012}; prokaryotes  \cite{Unwin2003};  and artificial, computer-generated,  DNA sequences.  This method may also complement  other taxonomic analyses by bringing in additional information gleaned from comparisons of  non-homologous  and non-coding sequences.

An example of the CGR/DSSIM/MDS approach is the Molecular Distance Map in Figure~\ref{vertebrates} which depicts the complete mitochondrial DNA sequences  of all 1,791  jawed vertebrates in our dataset. (In the legends of Figures 2-6,  the number of represented mtDNA sequences  in each category is listed in paranthesis after the category name.)  Note  that the position of each point in a map is determined by  {\it all}  the distances between the sequence it represents and the other sequences in the dataset. In  the case of Figure~\ref{vertebrates}, the position of each sequence-point is determined by the 1,790 numerical distances between its sequence and all the other mtDNA vertebrate sequences in that dataset.

Observe that all five different  subphyla of jawed vertebrates are separated in non-overlapping clusters, with very few exceptions. Examples of fish species bordering or slightly mixed with the amphibian cluster include {\it Polypterus ornatipinnis} (\#3125, ornate bichir), {\it Polypterus senegalus} (\#2868, Senegal bichir), both  with primitive pairs of lungs; {\it Erpetoichthys calabaricus} (\#2745, reedfish) who can breathe atmospheric air using a pair of lungs;  and {\it Porichtys myriaster} (\#2483, specklefish midshipman) a toadfish of the order Batrachoidiformes. It is noteworthy that the question of whether species of the {\it Polypterus} genus are fish or amphibians has been discussed extensively for hundreds of years \cite{Hall2001}. Interestingly, all four represented lungfish (a.k.a. salamanderfish), are  also bordering the amphibian cluster:  {\it Protopterus aethiopicus} (\#873, marbled lungfish), {\it Lepidosiren paradoxa} (\#2910, South American lungfish), {\it Neoceratodus forsteri} (\#2957, Australian lungfish), {\it Protopterus doloi} (\#3119, spotted African lungfish). Note that, in answer to the hypothesis in \cite{Edwards2002} regarding the diversity of signatures across vertebrates, in Figure \ref{vertebrates}, the  avian mtDNA signatures cluster neither with the mammals nor with the reptiles, and form a completely separate cluster of their own (albeit closer to reptiles than to mammals).

We applied our method to visualize the relationships among all   represented species from  the (super) kingdom Protista whose taxon, as defined in the legend of  Figure \ref{AllProtists},  had  more than one representative.  As expected,  the maximum distance between pairs of  sequences in this map was higher than the maximum distances for the other maps  in this paper, all at lower taxonomic levels.
  
The most obvious outlier in  Figure \ref{AllProtists}  is {\it Haemoproteus} sp. jb1.JA27 (\#1466), sequenced in \cite{Beadell2005} (see also \cite{Valkiunas2010}), and listed as an {\it unclassified} organism in the NCBI taxonomy.  Note first that  this species-point belongs to the same kingdom (Chromalveolata), superphylum (Alveolata), phylum (Apicomplexa), and class (Aconoidasida),  as the other two species-points  that appear grouped with it, {\it Babesia bovis} T2Bo  (\#1935), and  {\it Theileria parva} (\#3173).  This indicates that its position is not fully anomalous. Moreover,  as indicated by the high value of {\it Stress-1} for this figure, an inspection of  DSSIM distances  shows that this species-point may not be a true outlier, and its position may not be as striking in a higher dimensional version of the Molecular Distance  Map.  Overall,  this  map shows that  our method allows an exploration of diversity at the level of super kingdom,   obtains  good clustering of known subtaxonomic groups, while at the same time indicating  a lack of genome sequence information  and  paucity of representation that complicates analyses for this fascinating taxonomic group.

We then applied our method to visualize the relationships  between all available complete mtDNA sequences from three classes, Amphibia, Insecta and Mammalia (Figure \ref{insects_mammals_amphibians}), as well as  observe relationships  within class Amphibia and three of its orders  (Figure \ref{amphibians}).
     % and class Insecta grouped in  nine categories  in Figure \ref{insects}.
Note that a feature of  MDS  is that the points $p_{i}$ are not unique. Indeed, one can translate or rotate a  map without affecting the pairwise spatial distances $d(i, j) = ||p_{i}-p_{j}||$. In addition, the obtained points in an MDS map may change coordinates when more data items are added to or removed from the dataset. This is because the output of the MDS aims to preserve only the pairwise spatial distances between points, and this can be achieved even when some of the points change their coordinates. In particular, the  $(x, y)$-coordinates of a  point representing an amphibian species in the amphibians-insects-mammals map (Figure \ref{insects_mammals_amphibians}) will not necessarily be the same as the $(x, y)$-coordinates of the same  point when only amphibians are mapped (Figure \ref{amphibians}).

In general, Molecular Distance Maps are in  good agreement with classical phylogenetic trees at all scales of taxonomic comparisons,  see  Figure \ref{amphibians} with \cite{Pyron2011}, and Figure \ref{primatesonly} with\cite{Shoshani1996}. In addition, our approach may  be able to weigh in on conflicts between taxonomic classifications  based on morphological traits and those based on more  recent molecular data, as in the case of tarsiers, as seen below.

Zooming in, we observed the relationships within an order, Primates, with its  suborders (Figure \ref{primatesonly}). Notably, two extinct species of the genus {\it Homo}  are represented: {\it Homo sapiens neanderthalensis} and {\it Homo sapiens ssp. Denisova}. Primates can be classified into two groups, Haplorhini  (dry-nosed primates comprising anthropoids and tarsiers) and Strepsirrhini (wet-nosed primates including lemurs and lorises). The map shows a  clear separation of  these suborders, with the top-left arm of the map in Figure \ref{primatesonly}, comprising the Strepsirrhini. However, there are two  Haplorhini placed in the Strepsirrhini cluster, namely  {\it Tarsius bancanus} (\#2978, Horsfield's tarsier) and {\it Tarsius syrichta} (\#1381, Philippine tarsier).  The phylogenetic placement of tarsiers within the  order Primates has been  controversial  for over a century, \cite{Jameson2011}. According to \cite{Chatterjee2009}, mitochondrial DNA evidence places tarsiiformes as a sister group to Strepsirrhini, while in contrast, \cite{Perelman2011} places tarsiers within Happlorhini. In Figure \ref{primatesonly}  the tarsiers are located within the Strepsrrhini cluster, thus agreeing with \cite{Chatterjee2009}. This may be partly because both this study and \cite{Chatterjee2009} used mitochondrial DNA, whose signature may be different from that of chromosomal DNA as seen in Figure \ref{fig:CGR}(left) and Figure \ref{fig:CGR}(center).
 
The DSSIM distances computed between all pairs of complete mtDNA sequences varied in range. The minimum  distance  was 0, between two pairs of identical  mtDNA sequences. The first pair  comprised the mtDNA of {\it Rhinomugil nasutus} (\#98, shark mullet, length 16,974 bp) and {\it Moolgarda cunnesius} (\#103, longarm mullet,  length 16,974 bp).  A base-to-base sequence comparison between these sequences (\#98, NC\_017897.1; \#103, NC\_017902.1) showed  that the sequences were indeed identical. However, after completion of this work, the sequence for species \#103 was updated to a new version  (NC\_017902.2), on  7 March, 2013, and is now different from the sequence for species \#98 (NC\_017897.1). The second pair  comprises the mtDNA  sequences \#1033 and \#1034 (length 16,623 bp), generated by crossing female {\it Megalobrama amblycephala}  with  male {\it Xenocypris davidi}  leading to the creation of both diploid (\#1033) and triploid (\#1034) nuclear genomes, \cite{Hu2012}, but identical mitochondrial genomes.

The maximum distance was found to be between  {\it Pseudendoclonium akinetum} (\# 2656,  a green alga, length 95,880) and  {\it Candida subhashii} (\#954, a yeast,  length 29,795). Interestingly, the pair with the  maximum distance $\Delta (\# 2656, \# 954) = 1.0033$ featured neither the longest mitochondrial sequence,  with the darkest CGR  ({\it Cucumis sativus}, \#533, cucumber, length 1,555,935 bp),
 nor the shortest mitochondrial sequence, with the lightest CGR ({\it Silene conica},  \#440, sand catchfly, a plant,  length 288 bp).

An inspection  of the   distances between {\it Homo sapiens sapiens} and all the other primate mitochondrial genomes in the dataset showed that the minimum distance  to {\it Homo sapiens sapiens} was $\Delta (\# 1321, \# 1720)$ $= 0.1340$,  the distance to {\it Homo sapiens neanderthalensis} (\#1720, Neanderthal),  with the second smallest distance to it being  $\Delta(\# 1321, \# 1052) = 0.2280$, the distance to {\it Homo sapiens ssp. Denisova} (\#1052, Denisovan). The third smallest distance was $\Delta(\# 1321, \# 3084) =  0.5591$ to {\it Pan troglodytes} (\#3084, chimp).  Figure \ref{human_graph} shows the graph of the distances between the {\it Homo sapiens sapiens} mtDNA and each of the  primate mitochondrial genomes. With no exceptions, this graph is in full agreement with established phylogenetic trees, \cite{Shoshani1996}. The largest distance between  the {\it Homo sapiens sapiens}  mtDNA and another mtDNA sequence in the dataset was 0.9957, the distance  between {\it Homo sapiens sapiens} and  {\it Cucumis sativus} (\#533, cucumber, length 1,555,935 bp).

In addition to comparing real DNA sequences,  this method can compare  real DNA sequences to computer-generated sequences. As an example, we compared the mtDNA genome of {\it Homo sapiens sapiens} with one hundred artificial, computer-generated, DNA sequences of the same length and the  same  trinucleotide frequencies as the original. The average  distance between these artificial sequences and the original human mitochondrial DNA is 0.8991. This indicates that   all  ``human''  artificial DNA sequences are  more distant from the {\it Homo sapiens sapiens} mitochondrial genome than {\it Drosophila melanogaster } (\#3120, fruit fly) mtDNA, with $\Delta (\# 3120, \# 1321) = 0.8572$. This further implies that  trinucleotide frequencies may not contain sufficient  information to classify  a genetic sequence, suggesting that Goldman's claim \cite{Goldman1993} that ``CGR  gives no futher insight into the structure of the DNA sequence than is given by the  dinucleotide and trinucleotide frequencies'' may not hold in general.

The {\it Stress-1} values for all but one of the  Molecular Distance Maps in this paper were in the ``acceptable'' range $[0, 0.2]$. The exception was Figure \ref{AllProtists} with {\it Stress-1} equal to 0.26. Note that   {\it Stress-1} generally decreases with an increase in dimensionality, from $q = 2$ to $q = 3, 4, 5...$. Note also that, as suggested  in \cite{BorgGroenen2010}, the {\it Stress-1} guidelines are not absolute: It is not always the case that only MDS  representations with {\it Stress-1} under $0.2$ are acceptable, nor that all MDS representations with {\it Stress-1}  under $0.05$ are good.

In all the  calculations in this paper, we  used the full mitochondrial sequences. However, since  the length of a sequence can influence the brightness of its CGR and thus  its Molecular Distance Map coordinates, further analysis is needed to elucidate the effect of sequence length on  the positions  of sequence-points in a Molecular Distance Map.  The choice of length of DNA sequences used  may ultimately depend  on the particular dataset and particular application.

We now  discuss some limitations of the proposed methods. Firstly, DSSIM is  very effective  at picking up subtle differences between images. For example,  all vertebrate CGRs present  the triangular fractal structure seen in the human mtDNA, and are visually very similar, as seen in Figure \ref{fig:CGR}(left) and Figure  \ref{fig:CGR}(right).  In spite of this, DSSIM is able to detect a  range of differences that is sufficient for a good positioning  of all 1,791 mtDNA sequences relative to each other. This being said,  DSSIM may give too much weight to subtle differences, so that  small and big  differences in images  produce distances that are numerically very close.  This may be a useful feature for the analysis of datasets of closely related sequences. For large-scale taxonomic comparisons however, refinements of DSSIM or  the use of  other distances needs to be explored, that would space further apart the values of  distances arising from small differences versus  those arising from big-pattern differences between images.

Secondly, MDS  always has some errors, in the sense that the spatial distance between two points does not always reflect the original distance in the distance matrix.  For fine analyses,  the placement of a sequence-point in a map has to be confirmed by checking the original distance matrix. Possible solutions include increasing the dimensionality of the maps to three-dimensional maps, which  are still easily interpretable visually and have been shown in some cases to  separate clusters which seemed  incorrectly intermeshed in the two-dimensional version of the map. Other possibilities include a colour-scheme that would colour  points with low stress-per-point differently from  the ones with high stress-per-point, and  thus alert the user to the  regions where discrepancies between the spatial distance and the original distance exist.

 Thirdly,   we note  that the use of the particular distance measure (DSSIM) or particular scaling technique (classical MDS)  does not mean that these are the optimal  choices in all cases.
 
Lastly,  as seen in Figure  \ref{fig:CGR}(left) and Figure \ref{fig:CGR}(center), the genomic signature of mtDNA can be very  different from that of nuclear DNA of the same species and care must be employed in  choosing the dataset and interpreting the results.

\section*{Conclusions}
Our analysis suggests that the  oligomer composition   of mitochondrial  DNA sequences can be a source of taxonomic information. These results are of interest both because of the large dataset considered (see, e.g.,  the correct grouping in taxonomic categories of 1,791 mitochondrial genomes in Figure \ref{vertebrates}),  and  because this information is extracted from DNA sequences  that, by normal criteria, would not be considered homologous. 

Potential applications of Molecular Distance Maps - when used on a dataset of  genomic sequences, whether  coding or non-coding, homologous or not homologous, of the same length or vastly different lengths -- include  identification of large evolutionary lineages,  taxonomic classifications,  species identification, as well as possible quantitative  definitions of the notion of species and other taxa. 
 
 Possible extensions include generalizations of MDS, such as 3-dimensional MDS, for improved visualization, and the use of increased oligomer length  (higher values of $k$) for comparisons of longer subsequences  in case of whole chromosome or whole genome analyses. We  note also that this method can be applied to analyzing sequences over other alphabets. For example binary sequences could be imaged  using a square  with vertices  labelled 00, 01, 10, 11, and then  DSSIM and MDS could be employed to  compare and map them.

% Do NOT remove this, even if you are not including acknowledgments
\section*{Acknowledgments}
We thank Ronghai Tu for an early version of our MATLAB code to generate CGR images, Tao Tao for assistance with NCBI's GenBank,  Steffen Kopecki for generating artificial sequences and  discussions. We also thank Andre Lachance, Jeremy McNeill, and Greg Thorn for resources and discussions on taxonomy. We thank the Oxford University Mathematical Institute for the use of their Windows compute server Pootle/WTS. 

%This work was supported by  Natural Sciences and Engineering Research Council of Canada (NSERC) Discovery Grants to L.K. and K.H.; Oxford University Press Clarendon Fund and NSERC USRA, PGSM, and PGSD3 awards to N.D.; NSERC USRA award to N.B.

%\section*{References}
% The bibtex filename
\bibliography{MoDMaps}

\newpage

\section*{Figures}
\begin{figure}[!ht]
\begin{center}
       \subfigure{
%\CGRplot{human_mitochondrion_J01415.png}%
       }\hspace{.3cm}
        \subfigure{
%\CGRplot{human_chromosome11_betaGlobinRegion.png}
        }\hspace{.3cm}
       \subfigure{
%\CGRplot{polypterusEndlicherii_mitochondrion_NC_020791.png}
}
\end{center} 
\caption{
{\bf CGR images for  three  DNA sequences.  LEFT:}
 {\it Homo sapiens sapiens} mtDNA,  16,569 bp; {\bf CENTER:}
  {\it Homo sapiens sapiens} chromosome 11, beta-globin region, 73,308 bp; {\bf RIGHT:}
 {\it Polypterus endlicherii} (fish) mtDNA,  16,632 bp.
  % Prominent diagonals are indicative of purine (A/G) and pyrimidine (C/T) runs, sparsness in the upper half indicates low G+C content.
    Note that chromosomal and mitochondrial DNA from the same species can display different  patterns, and also that mtDNA of different species may display visually similar patterns that are however sufficiently different as to be computationally distinguishable. 
}
\label{fig:CGR}
\end{figure}

\begin{figure}[!ht]
\begin{center}
\subfigure{
} \\
\subfigure{
}
\end{center}
\caption{
{\bf TOP: Molecular Distance Map of 
 phylum Vertebrata (excluding the 5 represented jawless vertebrates), with its five subphyla.}  The total number of mtDNA sequences is 1,791, the average DSSIM distance is 0.8652,
        and  the MDS {\it Stress-1}  is 0.12. Fish species bordering amphibians include  fish with primitive pairs of lungs 
        ({\it Polypterus ornatipinnis} \#3125, {\it Polypterus senegalus} \#2868),  a fish  who can breathe atmospheric air using a pair of lungs ({\it Erpetoichthys calabaricus} \#2745), a  toadfish ({\it Porichtys myriaster} \#2483), and all four represented  lungfish   
        ({\it Protopterus aethiopicus} \#873,
{\it Lepidosiren paradoxa} \#2910,
{\it Neoceratodus forsteri} \#2957,
{\it Protopterus doloi} \#3119). Note that the question  of whether species of the genus {\it Polypterus}  are fish or
amphibians has been discussed extensively for hundreds of years.  Note that gaps and spaces in clusters, in this and other maps, may be due to sampling bias.  {\bf BOTTOM:}  Screenshot of the 
zoomed-in rectangular region outlined in the TOP map, obtained using the interactive web tool  {\it MoD-Map}  \cite{MoDMap}. 
}
\label{vertebrates}
\end{figure}

\begin{figure}[!ht]
\begin{center}
\end{center}
\caption{
{\bf Molecular Distance Map of all represented species  from  (super)kingdom 
Protista and its orders.}  The total number of mtDNA sequences is 70, the average DSSIM distance is 0.8288,
        and  the MDS {\it Stress-1}  is 0.26. The sequence-point \#1466 (red) is the unclassified  {\it Haemoproteus} sp. jb1.JA27, \#1935 (grey) is {\it Babesia bovis T2Bo},
and \#3173 (grey) is  {\it Theileria parva}. 
The annotation shows that all these three 
species belong to the same taxonomic groups, Chromalveolata, Alveolata, Apicomplexa, Aconoidasida, up to the order level.
}
\label{AllProtists} 
\end{figure}

\begin{figure}[!ht]
\begin{center}
\end{center}
\caption{
{\bf Molecular Distance Map of three classes: Amphibia, Insecta and Mammalia.}   Note that the method successfully clusters taxonomic groups also at the Class level. Gaps and spaces in clusters, in this and other maps, may be due to sampling bias.  A topic of further exploration would be to understand  the cluster shapes  and nature of the distribution of sequences in this figure. The total number of mtDNA sequences is 790, the average DSSIM distance is 0.8139, and the MDS {\it Stress-1} is 0.16.
}
\label{insects_mammals_amphibians}
\end{figure}

\begin{figure}[!ht]
\begin{center}
\end{center}
\caption{
{\bf Molecular Distance Map of Class Amphibia and  three of its  orders.}  The total number of mtDNA sequences is 112,   the average DSSIM distance is  0.8445,
 and the MDS {\it Stress-1} is 0.18. Note that the shape of the amphibian cluster and the $(x, y)$-coordinates of sequence-points are different here from those in Figure \ref{insects_mammals_amphibians}. This is because MDS outputs a map that aims to preserve pairwise distances between points, but not necessarily their absolute coordinates.
}
\label{amphibians} 
\end{figure}

\begin{figure}[!ht]
\begin{center}
\end{center}
\caption{
{\bf Order Primates and its suborders: Haplorhini (anthropoids and tarsiers), and 
 Strepsirrhini (lemurs and lorises).}  The total number of mtDNA sequences is 62, 
the average DSSIM distance is 0.7733,  and the MDS {\it Stress-1} is 0.19. The outliers are {\it Tarsius syrichta} \#1381, and {\it Tarsius bancanus} \#2978, whose  placement within the order Primates  has been subject of debate for over a century.
}
\label{primatesonly}
\end{figure}

\begin{figure}[!ht]
\begin{center}
\end{center}
\caption{
{\bf Graph of the DSSIM distances between the CGR images of {\it Homo sapiens sapiens} mtDNA 
and each of the 62 primate  mitochondrial genomes (sorted by their distance from the human mtDNA).}  The distances  are in accordance with  established phylogenetic trees: The species with the smallest DSSIM distances from  {\it Homo sapiens sapiens} are {\it Homo sapiens neanderthalensis},   {\it Home sapiens ssp. Denisova}, followed by  the   chimp.
}
\label{human_graph}
\end{figure}

\newpage
\newpage
\newpage
\newpage
For a complete version, including figures, please see\\ \url{http://www.csd.uwo.ca/~lila/pdfs/Space_of_Genomic_Signatures.pdf}

\end{document}